\documentclass[journal=jctcce,manuscript=article,layout=traditional]{achemso}

\usepackage{chemformula} % Formula subscripts using \ch{}
\usepackage[T1]{fontenc} % Use modern font encodings
\usepackage[colorlinks=true]{hyperref}
\usepackage{algorithm}
\usepackage{algorithmic}
\usepackage{multirow}
\usepackage{subfig}
\usepackage{booktabs}

\newcommand{\reffig}[1]{Figure~\ref{#1}}
\newcommand{\reftab}[1]{Table~\ref{#1}}
\newcommand{\refsec}[1]{Section~\ref{#1}}
\newcommand{\refequ}[1]{Equation~\ref{#1}}
\newcommand{\etal}{et~al.~}
\newcommand{\ang}{$\text{\AA}$}

\author{Satoshi Imamura}
\email{s-imamura@fujitsu.com}
\author{Akihiko Kasagi}
\author{Eiji Yoshida}
\affiliation[Fujitsu Limited]
{Computing Laboratory, Fujitsu Limited} 
\title{Accurate and Fast Geometry Optimization with Time Estimation and Method Switching}


\begin{document}

\begin{tocentry}
  \centering
  \includegraphics[width=8.25cm]{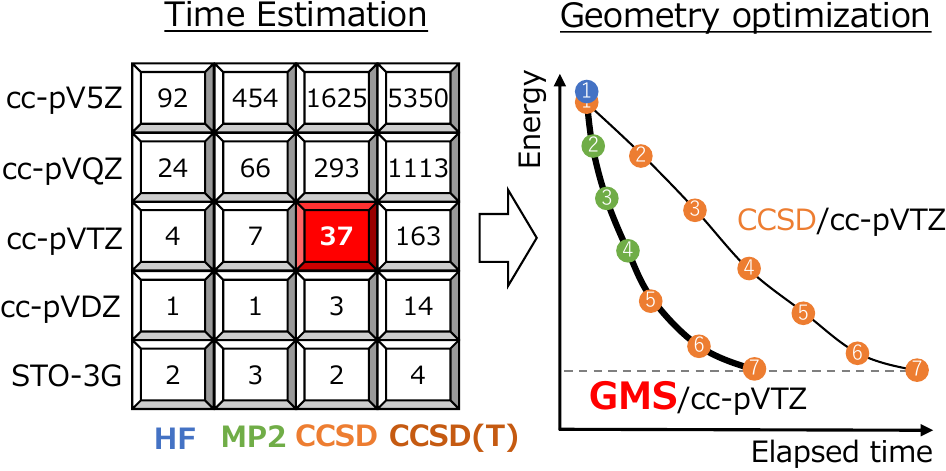}
\end{tocentry}

\begin{abstract}
Geometry optimization is an important task in quantum chemical calculations to analyze the characteristics of molecules. A top concern on it is a long execution time because time-consuming energy and gradient calculations are repeated across several to tens of steps. In this work, we present a scheme to estimate the execution times of geometry optimization of a target molecule at different accuracy levels (i.e., the combinations of ab initio methods and basis sets). It enables to identify the accuracy levels where geometry optimization will finish in an acceptable time. In addition, we propose a \emph{gradient-based method switching (GMS)} technique that reduces the execution time by dynamically switching multiple methods during geometry optimization. Our evaluation using 46 molecules in total shows that the geometry optimization times at 20 accuracy levels are estimated with a mean error of 29.5\%, and GMS reduces the execution time by up to 42.7\% without affecting the accuracy of geometry optimization.
\end{abstract}

\section{Introduction} \label{sec:introduction}

Geometry optimization is a process to find the atomic coordinates that minimize the energy of a target molecule. It is an important basis task in quantum chemical calculations because the optimized geometries are used to analyze the molecular characteristics and structures~\cite{Schlegel:2003ex, Schlegel:2011go, Shajan:2023ge}. In geometry optimization, a stationary point on a potential energy surface (PES) is explored by iteratively calculating the energy and gradients of a molecule while changing its atomic coordinates step by step.

With the Taylor series, the energy at a point $x$ on a PES, $E(x)$, is represented in a quadratic approximation with respect to a near point $x_0$,
\begin{equation*}
    E(x) = E(x_0) + \boldsymbol{G}^T(x_0) \Delta x + \frac{1}{2} \Delta x^T \boldsymbol{H}(x_0) \Delta x,
\end{equation*}
where $\boldsymbol{G}(x_0)$ is the gradient vector ($dE/dx$) at $x_0$, $\Delta x = x - x_0$, and $\boldsymbol{H}(x_0)$ is the Hessian matrix ($d^2E/dx^2$) at $x_0$. By differentiating the equation with respect to coordinates, the gradients at $x$, $\boldsymbol{G}(x)$, is represented in a quadratic approximation as
\begin{equation*}
    \boldsymbol{G}(x) = \boldsymbol{G}(x_0) + \boldsymbol{H}(x_0) \Delta x.
\end{equation*}
As $\boldsymbol{G}(x)$ becomes zero at a stationary point on a PES, the displacement to the stationary point, $\Delta x$, is given by
\begin{equation*}
    \Delta x = -\boldsymbol{H}(x_0)^{-1} \boldsymbol{G}(x_0).
\end{equation*}
Solving this equation is called the \emph{Newton-Raphson} step, which is a core part of geometry optimization. $\boldsymbol{G}(x_0)$ is obtained by differentiating $E(x_0)$ with respect to the coordinates. On the other hand, $\boldsymbol{H}(x_0)$, which is hard to calculate exactly, is commonly approximated with a quasi-Newton method such as the \emph{Broyden-Fletcher-Goldfarb-Shanno (BFGS)} method~\cite{Fischer:1992ge}.

In every step of geometry optimization, energy and gradient calculations at current coordinates are performed. For both of them, a wide variety of ab initio methods with different accuracy and computational costs are available, such as Hartree-Fock method (HF)~\cite{Baerends:1973se}, density function theory (DFT)~\cite{Kohn:1965:se}, M{\o}ller-Plesset perturbation theory (MP)~\cite{Moller:1934no}, configuration interaction theory (CI)~\cite{David:1999co}, and coupled cluster theory (CC)~\cite{Bartlett:2007co}. There is basically a trade-off between an accuracy and computational cost among them, which means that more accurate methods require higher computational costs. 

To improve the efficiency of geometry optimization, various approaches have been proposed. Chaudhuri and Freed extended the \emph{improved virtual orbital-complete active space configuration interaction (IVO-CASCI)} method to enable geometry optimization and vibrational frequency calculation~\cite{Chaudhuri:2007ge}. It achieved a comparable or higher accuracy compared to \emph{configuration interaction singles (CIS)} and \emph{complete active space self-consistent field (CASSCF)} with a lower computational cost. Park implemented the analytical gradient theory for the \emph{adaptive sampling CI SCF (ASCI-SCF)} method~\cite{Park:2021se}. It achieved a good accuracy with large active spaces by approximating gradients depending on the sampled determinants. Warden \etal examined several \emph{focal-point methods} combining MP methods with coupled cluster singles, doubles, and perturbative triples [CCSD(T)] to achieve a high accuracy with a lower computational cost~\cite{Warden:2020ef}. Sahu \etal enabled geometry optimization and vibrational spectra calculation for proteins by combining the \emph{molecular tailoring approach (MTA)} with DFT and utilizing large-scale parallelization on supercomputers~\cite{Sahu:2023co}. Khire \etal also applied MTA to enable the PES construction of medium-sized molecules at the CCSD(T)/aug-cc-pVTZ level~\cite{Khire:2022en}. Ahuja \etal applied a reinforcement learning approach that produces a correction term for the quasi-Newton step with BFGS to improve the convergence of geometry optimization~\cite{Ahuja:2021le}. Delgado \etal proposed a variational quantum algorithm to perform geometry optimization using a quantum computer~\cite{Delgado:2021va}. It minimizes a general cost function in a variational scheme by simultaneously optimizing both the \emph{ansatz} parameters and  Hamiltonian parameters. It achieved a good agreement to the full configuration interaction (FCI) method in a noise-less quantum computer simulation.

The in-depth evaluation of geometry optimization has also been conducted. Cremer \etal compared the accuracy of geometry optimization with several MP and CC methods within large correlation consistent basis sets~\cite{Cremer:2001ex}. Their evaluation showed that the CCSD(T)/cc-pVTZ and CCSD(T)/cc-pVQZ levels achieve a very high accuracy. B\'{a}lint and J\"{a}ntschi compared the 39 combinations of various methods and basis sets to analyze the relationship between them and to determine which to use under different circumstances~\cite{Balint:2021co}. Shajan \etal compared various open-source geometry optimization implementations via their open-source interface~\cite{Shajan:2023ge}. They demonstrated that \emph{internal coordinates}, which represent molecular structures with bond lengths, bond angles, and torsion angles, achieved the better convergence than Cartesian coordinates, and the choice of the initial Hessian and Hessian update method in quasi-Newton approaches also contribute to the convergence.

Recently, surrogate models that predict PESs at the DFT level have been studied intensively to reduce the computational cost of geometry optimization. R\'{i}o \etal~\cite{Rio:2019lo} and Yang \etal~\cite{Yang:2021ma} presented active learning methods with a Gaussian process regression (GPR) model and neural network (NN) model, respectively. In an active learning process, DFT is executed to calculate accurate energy and gradients when the model prediction uncertainty is high, and surrogate models are updated with the new data. Laghuvarapu \etal proposed a NN model that predicts a molecular energy as the sum of energy contributions from bonds, angles, non-bonds, and dihedrals~\cite{Laghuvarapu:2020ba}. Born and K\"{a}stner extended a GPR model to internal coordinates and demonstrated that the convergence of geometry optimization is improved compared to a GPR model based on Cartesian coordinates~\cite{Born:2021ge}.

A top concern on geometry optimization is a long execution time because time-consuming energy and gradient calculations are repeated across several to tens of steps. Even if a surrogate model as introduced above is used for geometry optimization, ab initio calculations are still necessary to collect training data and complement the model prediction uncertainty. The times required for energy and gradient calculations at each step depend on methods, basis sets, and the size of molecules. High accuracy levels (e.g., CCSD(T) with large basis sets) are generally preferred in various calculations, such as rotational constants, vibrational frequencies, and chemical reactions~\cite{Warden:2020ef, Khire:2022en}. However, geometry optimization at such a high accuracy level cannot finish in a practical time for medium- or large-sized molecules. When the various sizes of molecules are required to be optimized, it is too arduous to manually select a practical accuracy level for each of them.

In this work, we present a scheme to estimate the execution times of geometry optimization of a target molecule at different accuracy levels. It enables to identify the accuracy levels where the geometry optimization of a target molecule finishes in an acceptable time and select an appropriate level from them. For instance, \reftab{tab:est_times_benzene} showing the estimated times for benzene tells us that geometry optimization at the CCSD/cc-pVQZ level will finish in one night, whereas that at the CCSD(T)/cc-pV5Z level will take around five days. Our evaluation demonstrates that the execution times at 20 accuracy levels are estimated with a mean error of 29.5\% for 16 molecules used by Puzzarini \etal\cite{Baker:1993Te}, and an appropriate accuracy level can be selected for each of the various sizes of 30 molecules in Baker set~\cite{Baker:1993Te} in addition to the 16 molecules based on the estimated times and a target time.

\begin{table}[t]
    \caption{The estimated execution times of the geometry optimization of benzene at 20 accuracy levels.}
    \label{tab:est_times_benzene}
    \centering
    \begin{tabular}{r|rrrr}
        \toprule
         & HF & MP2 & CCSD & CCSD(T) \\
        \midrule
        cc-pV5Z & 1h & 20h & 91h & {\bf 114h} \\
        cc-pVQZ & 6m & 2h & {\bf 9h} & 72h \\
        cc-pVTZ & 50s & 10m & 34m & 9h \\
        cc-pVDZ & 24s & 42s & 2m & 55m \\
        STO-3G & 10s & 5s & 27s & 1m \\
        \bottomrule
    \end{tabular}
\end{table}

In addition, we propose a dynamic method switching technique to reduce the execution time of geometry optimization. It uses light-weight methods at a first few steps and then appropriately switches to more accurate methods for the following steps, based on the norms of gradients obtained from the pre-executed geometry optimization at the lowest accuracy level (e.g., HF/STO-3G). Our evaluation shows that it reduces the execution time by a geometric mean of 22.2\% (up to 42.7\%) across 16 molecules in the Puzzarini set without any influence on the accuracy.

\newpage
\section{Methods} \label{sec:methodology}

\begin{table*}[t]
    \caption{Four molecule sets}
    \label{tab:molecule_suites}
    \centering
    \begin{tabular}{c|l}
        \toprule
        \bf{Molecule Set} & \multicolumn{1}{c}{\bf{Molecules}} \\
        \toprule
        \emph{Alkane} (10) & C$_n$H$_{2n+2}$ ($n = 1, 2, ..., 8, 10, 12$) \\
        \midrule
        \multirow{2}{*}{\emph{Small} (18)} & LiH, O$_2$, N$_2$, H$_2$O, BeH$_2$, NH$_3$, CO$_2$, HCl, CH$_4$, C$_2$H$_2$, C$_2$H$_4$, C$_2$H$_6$, C$_3$H$_4$, \\
         & C$_3$H$_6$, C$_3$H$_8$, C$_4$H$_6$, C$_4$H$_8$, C$_4$H$_{10}$ \\
        \midrule
        \multirow{6}{*}{\emph{Baker}~\cite{Baker:1993Te} (30)} & water, ammonia, ethane, acetylene, allene, hydroxysulfane, benzene, \\
        & methylamine, ethanol, acetone, disilyl-ether, 1,3,5-trisilacyclohexane, \\
        & benzaldehyde, 1,3-difluorobenzene, 1,3,5-trifluorobenzene, neopentane, furan, \\
        & naphthalene, 1,5-difluoronaphthalene, 2-hydroxybicyclopentane, ACHTAR10, \\ 
        & ACANIL01, benzidine, pterin, difuropyrazine, mesityl-oxide, histidine, \\
        & dimethylpentane, caffeine, menthone \\
        \midrule
        \multirow{2}{*}{\emph{Puzzarini}~\cite{Puzzarini:2008ac} (16)} & HF, N$_2$, CO, F$_2$, H$_2$O, HCN, HNC, CO$_2$, NH$_3$, CH$_4$, C$_2$H$_2$, HOF, HNO, \\
        & N$_2$H$_2$, C$_2$H$_4$, H$_2$CO \\
        \bottomrule
    \end{tabular}
\end{table*}

\subsection{Time Estimation} \label{subsec:time_estimation}

The execution time of geometry optimization, $T_{go}$, is represented as
\begin{equation}
    T_{go} = (T_e + T_g) \times S,
    \label{equ:time_go}
\end{equation}
where $T_e$ is an energy calculation time, $T_g$ is a gradient calculation time, and $S$ is the number of optimization steps. Hence, the estimation of $T_e$, $T_g$, and $S$ is necessary to estimate $T_{go}$.

\subsubsection{Estimation of $T_e$ and $T_g$}

The computational costs of ab initio methods basically depend on the number of basis functions, $N$. For instance, the general computational costs of HF, MP2, CCSD, and CCSD(T) are $O(N^3)$, $O(N^5)$, $O(N^6)$, and $O(N^7)$, respectively~\cite{Yoshioka:2021so}. However, the actual $T_e$ and $T_g$ of each method depend on its implementation and a machine configuration where it is executed. Therefore, to estimate $T_e$ and $T_g$, we use a linear regression model represented as
\begin{equation}
    log_{10}(T_{est}) = m \cdot log_{10}(N) + c,
    \label{equ:time_model}
\end{equation}
where $m$ is a regression coefficient corresponding to the exponent part of a computational cost $O(N^m)$, and $c$ is an intercept. The same model is used to estimate both $T_e$ and $T_g$. Note that the $T_e$ and $T_g$ of CCSD and CCSD(T) also strongly depend on the number of iterations in CCSD calculation; thus, the above model estimates the time taken per iteration for CCSD and CCSD(T).

In this work, we target four methods (HF, MP2, CCSD, and CCSD(T)) and five basis sets (STO-3G, cc-pVDZ, cc-pVTZ, cc-pVQZ, and cc-pV5Z) implemented in PySCF~\cite{pyscf}. An energy time model and gradient time model are fitted for each of the 20 accuracy levels using the execution times measured on our server (see \refsec{subsec:setup} for our experimental setup). $T_e$ and $T_g$ with STO-3G are measured for ten alkane molecules listed in \reftab{tab:molecule_suites}, while those with cc-pV\{D,T,Q,5\}Z are measured for 18 \emph{small} molecules. This model fitting procedure should be available for other methods and basis sets.

\reftab{tab:time_models} shows $m$, $c$, and the coefficient of determination, $R^2$, of the fitted energy and gradient time models. We can see that almost all the models are well fitted with a $R^2$ of over 0.75. The exceptions are the energy time models for HF/cc-pVTZ and MP2/cc-pVTZ, and the gradient time models for CCSD(T)/cc-pV\{Q,5\}Z. The former two cases are due to a sudden increase in $T_e$ for HF and MP2 with around 250 basis functions. In the latter two cases, the $T_g$ of CCSD(T) does not scale well to the number of basis functions due to the non-optimized implementation in PySCF~\cite{pyscf_slow_ccsd_t}. Moreover, we can also see that $m$ is basically larger for higher accuracy levels. Since $m$ corresponds to the exponent part of computational cost, $O(N^m)$, this observation is in a good agreement with the computational costs of the four methods.

\begin{table*}[t]
    \caption{The coefficients and $R^2$ of time estimation models.}
    \label{tab:time_models}
    \centering
    {\bf (a) Energy time models}
    \begin{tabular}{c|ccc|ccc|ccc|ccc}
        \toprule
        & \multicolumn{3}{c|}{\bf{HF}} & \multicolumn{3}{|c|}{{\bf MP2}} & \multicolumn{3}{|c|}{{\bf CCSD}} & \multicolumn{3}{|c}{{\bf CCSD(T)}} \\
        \toprule      
        \bf{Basis set} & $m$ & $c$ & $R^2$ & $m$ & $c$ & $R^2$ & $m$ & $c$ & $R^2$ & $m$ & $c$ & $R^2$ \\
        \midrule
        STO-3G & 0.67 & -1.19 & 0.91 & 0.60 & -1.03 & 0.90 & 1.40 & -2.55 & 0.91 & 1.03 & -2.04 & 0.96 \\
        cc-pVDZ & 0.84 & -1.48 & 0.90 & 0.77 & -1.29 & 0.87 & 1.21 & -2.43 & 0.79 & 1.60 & -2.99 & 0.96 \\
        cc-pVTZ & 1.10 & -2.00 & {\bf 0.70} & 1.21 & -2.14 & {\bf 0.75} & 2.49 & -4.95 & 0.95 & 2.54 & -4.78 & 0.98 \\
        cc-pVQZ & 2.40 & -4.80 & 0.92 & 2.66 & -5.28 & 0.93 & 3.60 & -7.48 & 0.98 & 3.04 & -5.79 & 0.98 \\
        cc-pV5Z & 3.27 & -6.88 & 0.97 & 3.60 & -7.52 & 0.97 & 4.27 & -9.12 & 0.98 & 3.77 & -7.47 & 0.99 \\
        \bottomrule
    \end{tabular}

    \begin{tabular}{c}
        \\
    \end{tabular}

    {\bf (b) Gradient time models}
    \begin{tabular}{c|ccc|ccc|ccc|ccc}
        \toprule
        & \multicolumn{3}{c|}{\bf{HF}} & \multicolumn{3}{|c|}{{\bf MP2}} & \multicolumn{3}{|c|}{{\bf CCSD}} & \multicolumn{3}{|c}{{\bf CCSD(T)}} \\
        \toprule      
        \bf{Basis set} & $m$ & $c$ & $R^2$ & $m$ & $c$ & $R^2$ & $m$ & $c$ & $R^2$ & $m$ & $c$ & $R^2$ \\
        \midrule
        STO-3G & 2.31 & -3.37 & 0.97 & 2.75 & -4.58 & 0.99 & 2.03 & -3.74 & 0.96 & 3.54 & -5.39 & 0.94 \\
        cc-pVDZ & 2.26 & -4.02 & 0.97 & 3.21 & -5.68 & 0.90 & 1.98 & -3.82 & 0.88 & 4.49 & -7.32 & 0.98 \\
        cc-pVTZ & 1.75 & -3.32 & 0.97 & 3.63 & -6.62 & 0.99 & 3.29 & -6.39 & 0.99 & 4.52 & -8.02 & 0.88 \\
        cc-pVQZ & 2.03 & -3.90 & 0.96 & 3.65 & -6.62 & 1.00 & 3.83 & -7.59 & 1.00 & 4.45 & -8.24 & {\bf 0.75} \\
        cc-pV5Z & 2.83 & -5.62 & 0.98 & 3.75 & -6.81 & 1.00 & 4.01 & -8.06 & 1.00 & 3.29 & -5.89 & {\bf 0.56} \\
        \bottomrule
    \end{tabular}
\end{table*}

To estimate the $T_e$ and $T_g$ of CCSD and CCSD(T), the number of iterations in CCSD calculation is necessary in addition to the time taken per iteration estimated with the regression models. Under the assumption that the number of iterations does not differ significantly with different basis sets, the number of iterations is obtained from the pre-executed energy calculation at the CCSD/STO-3G level for a target molecule. The time overhead of this pre-execution is negligible compared to geometry optimization with a higher accuracy level, because the energy calculation at the CCSD/STO-3G level is performed only once. For instance, the energy calculation at the CCSD/STO-3G level takes only three seconds for benzene, whereas the geometry optimization at the CCSD/cc-pVDZ level takes 257 seconds.

\subsubsection{Estimation of $S$} \label{subsubsec:estimation_S}

\begin{figure}[t]
    \centering
    \includegraphics[width=0.6\linewidth]{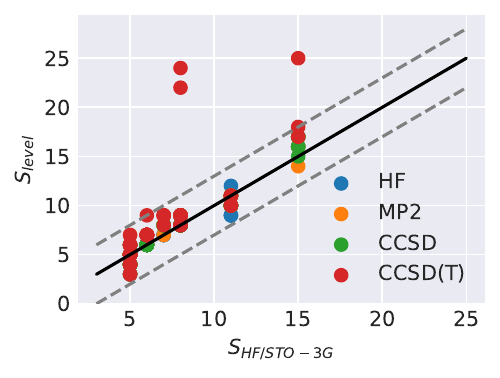}
    \caption{The number of steps at the HF/STO-3G level, $S_{HF/STO-3G}$, versus that at each of the 20 accuracy levels, $S_{level}$, for 16 molecules in Puzzarini set. The dotted gray lines show the region where the difference between $S_{HF/STO-3G}$ and $S_{level}$ is within three.}
    \label{fig:num_steps}
\end{figure}

The number of steps at an accuracy level, $S_{level}$, is estimated with the number of steps obtained from the pre-executed geometry optimization at the HF/STO-3G level, $S_{HF/STO-3G}$. \reffig{fig:num_steps} plots $S_{HF/STO-3G}$ versus $S_{level}$ at each of the 20 accuracy levels for 16 molecules in Puzzarini set listed in \reftab{tab:molecule_suites}. Note that the results at the CCSD(T)/cc-pV\{Q,5\}Z levels for the molecules with more than three atoms are not included, because geometry optimization does not finish in a practical time. We can see that $S_{HF/STO-3G}$ is a good estimator of $S_{level}$ because the difference between them is within three in almost all the results. There are only three exceptions out of 308 results: CCSD(T)/cc-pV\{D,T\}Z for HOF and CCSD(T)/cc-pVTZ for HNO. The time overhead of the pre-executed geometry optimization at the HF/STO-3G level is negligible compared to geometry optimization at a higher accuracy level. For instance, the geometry optimization of benzene at the HF/STO-3G level takes only 13 seconds, while that at the MP2/cc-pVTZ level takes 556 seconds.

\subsection{Gradient-based Method Switching (GMS)} \label{subsec:method_switching}

To reduce $T_{go}$ at a selected accuracy level, we propose a novel technique that dynamically switches multiple ab initio methods during geometry optimization. Its main concept is to save time by using light-weight methods at a few first steps where the selected accuracy level is unnecessary for energy and gradient calculations.

\begin{figure}[t]
    \centering
    \includegraphics[width=0.6\linewidth]{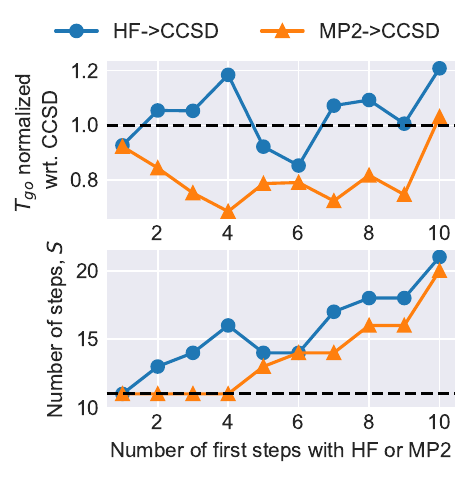}
    \caption{The geometry optimization time, $T_{go}$, normalized with respect to CCSD (upper) and the number of steps, $S$, (lower) when HF or MP2 is used at the $x$ first steps before switching to CCSD for caffeine with STO-3G.}
    \label{fig:dual_method_evaluation}
\end{figure}

We investigate how $T_{go}$ is affected by using light-weight methods at a few first steps. With the assumption that the CCSD/STO-3G level is selected, \reffig{fig:dual_method_evaluation} shows $T_{go}$ normalized with respect to CCSD and the number of steps, $S$, when HF or MP2 is used at a few first steps for caffeine/STO-3G. The x-axis indicates the number of first steps where HF or MP2 is used before switching to CCSD. When HF is used before CCSD ({\tt HF->CCSD}), $T_{go}$ is reduced by using HF only at the first step. Otherwise, $T_{go}$ is increased due to the increase of $S$. On the other hand, when MP2 is used before CCSD ({\tt MP2->CCSD}), $T_{go}$ is minimized without the increase of $S$ by using MP2 at the first four steps. From these results, we obtain the following two observations: (1) $T_{go}$ can be reduced by using light-weight methods at the appropriate number of first steps. (2) The appropriate number of first steps is different depending on light-weight methods.

\begin{figure}[t]
    \centering
    \includegraphics[width=0.6\linewidth]{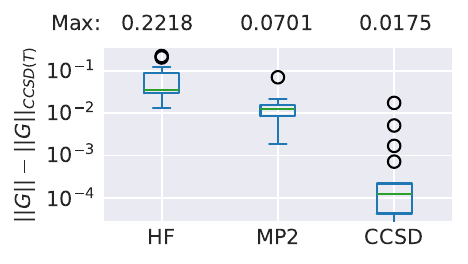}
    \caption{The error in the norm of gradients, $||\boldsymbol{G}||$, from that calculated with CCSD(T) for 18 small molecules with STO-3G.}
    \label{fig:norm_grad_error}
\end{figure}

To identify the appropriate number of first steps using light-weight methods, we focus on the norm of gradients, $||\boldsymbol{G}||$, calculated at each optimization step. It is a useful metric to know the calculation accuracy required at each step for two reasons: $||\boldsymbol{G}||$ can be calculated from the first step, and it decreases gradually as atomic coordinates get closer to the stationary ones. Hence, we evaluate the accuracy of $||\boldsymbol{G}||$ calculation with HF, MP2, and CCSD by comparing with CCSD(T). \reffig{fig:norm_grad_error} plots the error in $||\boldsymbol{G}||$ from that calculated with CCSD(T), $||\boldsymbol{G}||_{CCSD(T)}$, for 18 small molecules listed in \reftab{tab:molecule_suites} with STO-3G. We can see that more accurate methods generally achieve lower errors. Therefore, we use the maximum error of each method shown above the graph as a threshold to use it during geometry optimization.

\begin{figure}[t]
    \centering
    \includegraphics[width=0.6\linewidth]{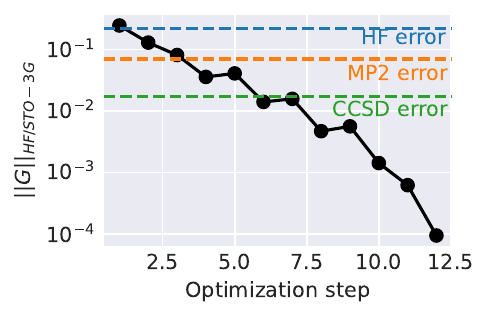}
    \caption{The norm of gradients calculated at the HF/STO-3G level, $||\boldsymbol{G}||_{HF/STO-3G}$, for caffeine.}
    \label{fig:norm_grad_caffeine}
\end{figure}

$||\boldsymbol{G}||$ at each step is obtained from the pre-executed geometry optimization of a target molecule at the HF/STO-3G level. \reffig{fig:norm_grad_caffeine} shows $||\boldsymbol{G}||$ calculated at each step at the HF/STO-3G level, $||\boldsymbol{G}||_{HF/STO-3G}$, for caffeine. The maximum errors in $||\boldsymbol{G}||$ calculation of HF, MP2, and CCSD evaluated in \reffig{fig:norm_grad_error} are shown with horizontal dotted lines. We implement the gradient-based method switching (GMS) technique that selects a method used at each step by checking whether $||\boldsymbol{G}||_{HF/STO-3G}$ exceeds the corresponding maximum error. For instance, when CCSD(T) with an arbitrary basis set is selected as an accuracy level for caffeine, a method at each step is selected as [HF, MP2, MP2, CCSD, CCSD, CCSD(T), ...]. As discussed in \refsec{subsubsec:estimation_S}, the time overhead of the pre-executed geometry optimization at the HF/STO-3G level is negligible compared to that with a higher accuracy level.

\subsection{Whole Procedure}

In this section, we summarize the whole procedure to estimate $T_{go}$ for a target molecule and perform geometry optimization at a selected accuracy level with our proposed GMS technique.

\subsubsection{Advance preparation}

The following two steps are required to be performed only once in advance for an experimental setup.

{\bf (a) Data collection:} First, $T_e$ and $T_g$ at all accuracy levels are measured for the molecule sets listed in \reftab{tab:molecule_suites} as learning data for the time estimation models. The ten alkane molecules and 18 small molecules are used for STO-3G and the other larger basis sets, respectively. The Cartesian coordinates of all the 28 molecules optimized with \emph{composite/CBS-Q} are obtained from CCCBDB~\cite{CCCBDB}. Second, the maximum errors in $||\boldsymbol{G}||$ calculation of all methods are evaluated for the 18 small molecules with STO-3G, as shown in \reffig{fig:norm_grad_error}. The values of $||\boldsymbol{G}||_{CCSD(T)}$ used as baselines are listed in the Supporting Information. In this work using the four methods and five basis sets, the whole data collection takes 31 hours in total. 
%If the experimental setup is similar to ours described in \refsec{subsec:setup} (i.e., the four methods implemented in PySCF are executed on a two-socket Xeon machine), one can skip the data collection and reuse our fitted models shown in \reftab{tab:time_models} and the maximum $||G||$ calculation errors shown in \reffig{fig:norm_grad_error}.

{\bf (b) Time estimation model fitting:} With $T_e$ and $T_g$ measured in the step (a) and the numbers of basis functions, $N$, of the 28 molecules, the linear regression models shown in \refequ{equ:time_model} are fitted to estimate $T_e$ and $T_g$ at all accuracy levels, as shown in \reftab{tab:time_models}.

\subsubsection{Geoemetry optimization of a target molecule}

{\bf (1) Pre-executions:} For the $T_{go}$ estimation and GMS, two pre-executions are necessary for a target molecule. First, the number of steps, $S_{HF/STO-3G}$, and the norm of gradients at each step, $||\boldsymbol{G}||_{HF/STO-3G}$, are obtained from the geometry optimization at the HF/STO-3G level. Second, if CCSD or CCSD(T) is included in target methods, the number of iterations in CCSD calculation is obtained from the energy calculation at the CCSD/STO-3G level. For benzidine which is the largest in Baker set, the geometry optimization at the HF/STO-3G level takes 154 seconds, and the energy calculation at the CCSD/STO-3G level takes 30 seconds.

{\bf (2) Time estimation and accuracy level selection:} $T_e$ and $T_g$ at all accuracy levels are estimated with the energy and gradient time models fitted in the step (b), the number of basis functions, $N$, of the target molecule, and the number of CCSD iterations obtained in the step (1). Then, $T_{go}$ at all accuracy levels are calculated based on \refequ{equ:time_go} with $S_{HF/STO-3G}$ obtained in the step (1) and the estimated $T_e$ and $T_g$. After that, an accuracy level where the estimated $T_{go}$ is acceptable can be selected.

{\bf (3) Geometry optimization with GMS:} The geometry optimization of the target molecule is performed with GMS at the accuracy level selected in the step (2). GMS selects a method used in each step by comparing $||\boldsymbol{G}||_{HF/STO-3G}$ obtained in the step (1) and the maximum errors evaluated in the step (a).

\section{Results and Discussion} \label{sec:evaluation}

In this section, we evaluate the estimation accuracy of $T_{go}$, the effectiveness of selecting an accuracy level based on the estimated $T_{go}$, and the time reduction by GMS. We first describe our experimental setup and then show the evaluation results.

\subsection{Experimental Setup} \label{subsec:setup}

In this work, we target 20 accuracy levels composed of four ab initio methods (HF, MP2, CCSD, and CCSD(T)) and five basis sets (STO-3G, cc-pVDZ, cc-pVTZ, cc-pVQZ, and cc-pV5Z) implemented in \emph{PySCF}~\cite{pyscf}. PySCF is a Python-based open-source quantum chemical calculation framework. We use the \emph{geomopt} module in PySCF via an interface to \emph{geomeTRIC}~\cite{geomeTRIC} with the default convergence criteria. The \emph{LinearRegression} module in \emph{scikit-learn}~\cite{scikit-learn} is used to fit the energy and gradient time estimation models. A server containing two Xeon Gold 6240M processors and 384~GB DRAM is used for all experiments in this work.

For evaluation, we select 16 molecules used by Puzzarini \etal\cite{Puzzarini:2008ac} and 30 molecules in Baker set~\cite{Baker:1993Te}, as listed in \reftab{tab:molecule_suites}. For the molecules in Puzzarini set, we obtain the Cartesian coordinates optimized with \emph{composite/CBS-Q} from CCCBDB~\cite{CCCBDB} and initialize all the bond distances to 1.0~\ang\ while keeping bond angles. Moreover, the experimental Cartesian coordinates of the molecules in Puzzarini set are also obtained from CCCBDB and used to evaluate the accuracy of optimized coordinates. For the molecules in Baker set, we use the initial Cartesian coordinates provided by Shajan \etal\cite{Shajan:2023ge}. 

\subsection{Time Estimation Accuracy} \label{subsec:model_accuracy}

\begin{figure}[t]
    \centering
    \includegraphics[width=0.6\linewidth]{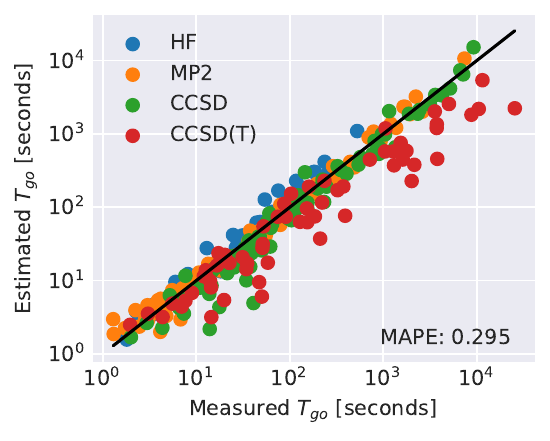}
    \caption{The measured versus estimated $T_{go}$ at the 20 accuracy levels for 16 molecules in Puzzarini set.}
    \label{fig:model_accuracy_all_basis}
\end{figure}

First, we evaluate the estimation accuracy of $T_{go}$ with \reffig{fig:model_accuracy_all_basis} plotting the measured versus estimated $T_{go}$ at the 20 accuracy levels for the 16 molecules in Puzzarini set. Different colored dots show the results of different methods, and the black line indicates the exact match between the estimated and measured $T_{go}$. Note that the results at the CCSD(T)/cc-pV\{Q,5\}Z levels for the molecules with more than three atoms are not included in similar to \reffig{fig:num_steps}. We can see that $T_{go}$ is estimated accurately in the most results. The mean absolute percentage error (MAPE) across all the results is 29.5\%, which is sufficiently low to identify the accuracy levels where geometry optimization finishes in an acceptable time. However, $T_{go}$ with CCSD(T) shown with red dots are under-estimated significantly in a lot of cases due to the low $R^2$ of the gradient time models for cc-pV\{Q,5\}Z as shown in \reftab{tab:time_models}. This is because the gradient calculation time of CCSD(T) does not scale well to the number of basis functions due to the non-optimized implementation in PySCF~\cite{pyscf_slow_ccsd_t}. It is our future work to optimize it or to consider a better time estimation model for the gradient calculation with CCSD(T).

\subsection{Accuracy Level Selection and Method Switching}

\begin{figure}[t]
    \centering    
    \includegraphics[width=\linewidth]{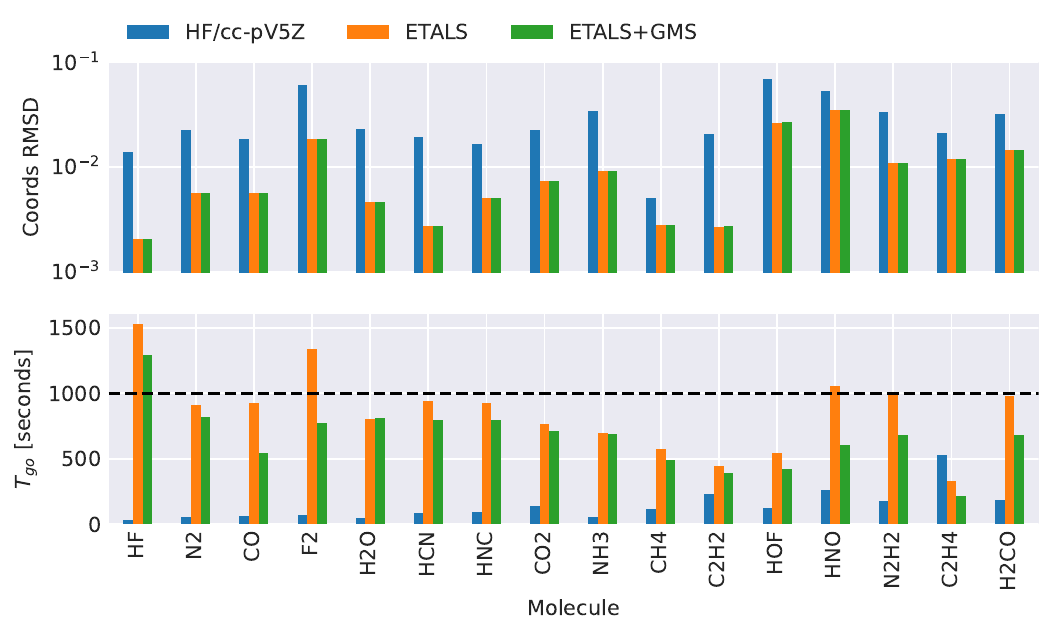}
    \caption{The RMSD of the optimized coordinates in \ang\ and the execution time of geometry optimization, $T_{go}$, for 16 molecules in Puzzarini set with a target $T_{go}$ of 1,000 seconds.}
    \label{fig:rmsd_time_test_mols_const1000}
\end{figure}

Next, we evaluate the effectiveness of the estimated time-based accuracy level selection (ETALS) and gradient-based method switching (GMS) technique. We set a target $T_{go}$ and select the highest accuracy level where the estimated $T_{go}$ is below it in the following three steps. (1) Without HF and STO-3G, the highest accuracy level meeting the target $T_{go}$ is searched with the outer loop in the order of larger basis sets and the inner loop in the order of more accurate methods. (2) If the accuracy level is not found in the step 1, the largest basis set with HF meeting the target $T_{go}$ is searched. (3) If the accuracy level is not found in the step 2, the most accurate method with STO-3G meeting the target $T_{go}$ is searched. For instance, when $T_{go}$ are estimated for benzene as shown in \reftab{tab:est_times_benzene} and a target $T_{go}$ is set to 300 seconds (5 minutes), CCSD/cc-pVDZ is selected.

\begin{figure}[t]
    \centering
    \subfloat[F2/cc-pV5Z]{\includegraphics[width=0.50\linewidth]{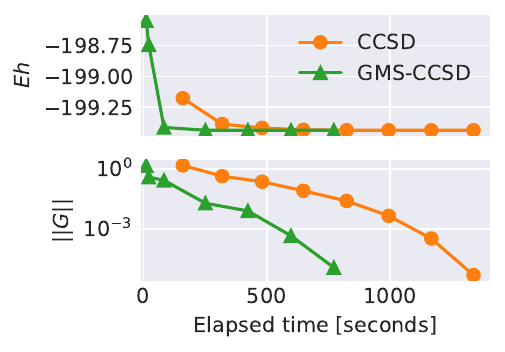}
    \label{subfig:geomopt_F2_ccpv5z}}
    \subfloat[Acetylene/cc-pVQZ]{\includegraphics[width=0.48\linewidth]{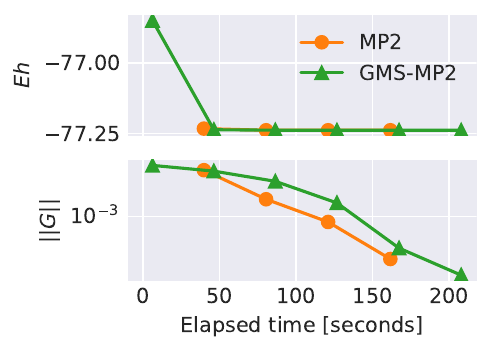}
    \label{subfig:geomopt_Acetylene_mp2_ccpvqz}}
    \caption{Hartree energy, $Eh$, (upper) and the norm of gradients, $||\boldsymbol{G}||$, (lower) calculated during geometry optimization. The x-axis shows the elapsed time in seconds.}
    \label{fig:geomopt}
\end{figure}

\begin{figure*}[t]
    \centering
    \includegraphics[width=\linewidth]{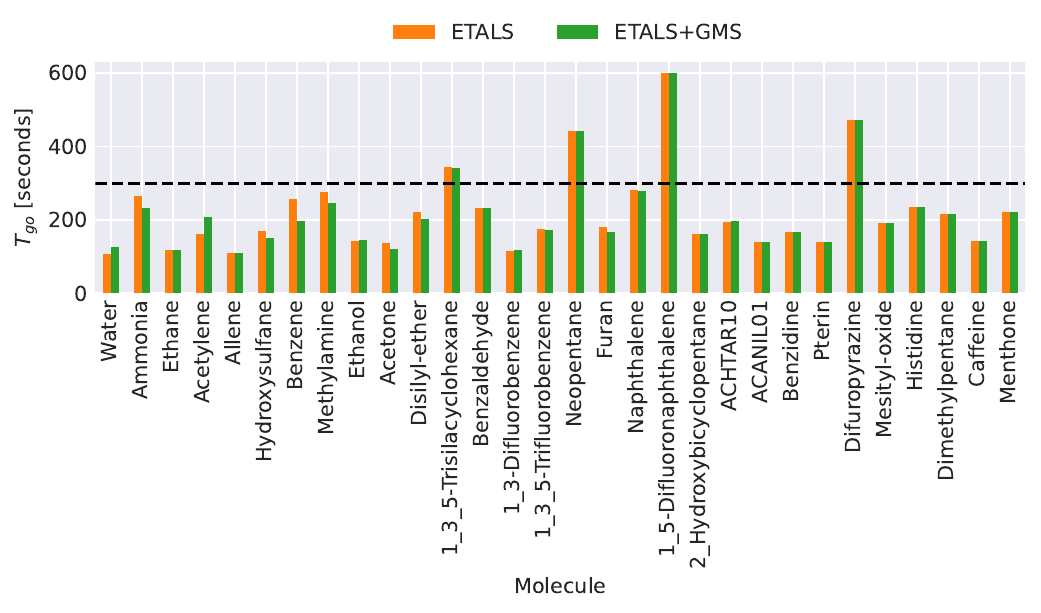}
    \caption{The execution time of geometry optimization, $T_{go}$, for 30 molecules in Baker set with a target $T_{go}$ of 300 seconds.}
    \label{fig:time_baker_mols_const300}
\end{figure*}

\reffig{fig:rmsd_time_test_mols_const1000} shows the root mean square deviation (RMSD) of the optimized coordinates with respect to the experimental coordinates and $T_{go}$ for the 16 molecules in Puzzarini set with a target $T_{go}$ of 1,000 seconds. The HF/cc-pV5Z level is evaluated as a naive baseline (blue bars), where the geometry optimization of all the molecules finishes in the target $T_{go}$. The orange bars show that the accuracy levels selected with ETALS achieve the much lower coordinates RMSD than HF/cc-pV5Z in around 1,000 seconds for almost all the molecules. This result demonstrates that ETALS enables to select a high accuracy level for each molecule based on the estimated and target $T_{go}$. 
%The accuracy levels selected for all the molecules are listed in the Supporting Information. 
The reason why $T_{go}$ exceeds the target $T_{go}$ by 50\% for the HF molecule is because the CCSD(T)/cc-pV5Z level is selected based on the under-estimated $T_{go}$ due to the low $R^2$ of the gradient time model as shown in \reftab{tab:time_models}. In addition, the green bars show that GMS reduces $T_{go}$ for almost all the molecules at the accuracy levels selected with ETALS without any influence on the coordinates RMSD. The maximum reduction is 42.7\% for F2, where CCSD/cc-pV5Z is selected. \reffig{subfig:geomopt_F2_ccpv5z} compares the energy and $||\boldsymbol{G}||$ calculated during the geometry optimization of F2 at the CCSD/cc-pV5Z level between without and with GMS. The x-axis shows the elapsed time in seconds, and the dots represent optimization steps. We can see that GMS reduces the number of steps using CCSD, where the time per step is around 160 seconds, from eight to four by using HF at the first two steps and MP2 at the third step. Although the energies calculated with HF at the first two steps are significantly different from those with CCSD, the final energy and $||\boldsymbol{G}||$ converge to the comparable values. The geometric mean of the $T_{go}$ reduction by GMS across the 16 molecules is 22.2\%. This is not a drastic reduction, but GMS can be applied without any concern because it does not affect the accuracy of geometry optimization. We also conduct the same evaluation with target $T_{go}$ of 100 and 300 seconds, where the geometric means of the $T_{go}$ reduction are 10.3\% and 15.8\%, respectively. 
%These results are included in the Supporting Information.

\reffig{fig:time_baker_mols_const300} plots $T_{go}$ for the 30 molecules in Baker set with a target $T_{go}$ of 300 seconds. The evaluation of coordinates RMSD is excluded because the experimental coordinates of almost all the molecules are not available. This graph shows that the appropriate accuracy levels are selected with ETALS so that geometry optimization finishes in around 300 seconds for almost all the molecules. As the various sizes of molecules are included in Baker set, the selected accuracy levels differ significantly depending on the molecular sizes. For instance, CCSD/cc-pVQZ is selected for water including three atoms, while HF/STO-3G is selected for menthone including 29 atoms.
%The accuracy levels selected for all the molecules are listed in the Supporting Information. 
For neopentane, 1\_5-difluoronaphthalene, and difuropyrazine, $T_{go}$ are relatively long because CCSD/cc-pVDZ and HF/cc-pVTZ are selected based on the under-estimated $T_{go}$ due to the low $R^2$ of the corresponding energy time models as shown in \reftab{tab:time_models}. GMS reduces $T_{go}$ for several small molecules at left side, where CCSD is mainly selected with ETALS, and the time is saved by using HF and MP2 at a few first steps. In contrast, the time reduction by GMS cannot be seen for large molecules at right side, because HF and MP2 are selected with ETALS. Unfortunately, GMS increases $T_{go}$ for water and acetylene because the number of steps is increased by using HF or MP2 at a few first steps. \reffig{subfig:geomopt_Acetylene_mp2_ccpvqz} shows the behavior of the geometry optimization of acetylene, where MP2/cc-pVQZ is selected with ETALS. While the number of steps is four without GMS, it is increased to six by using HF at the first step with GMS. Although the time increases by GMS can be avoided by conservatively setting higher thresholds to select light-weight methods, it would also decrease the time reduction by GMS. With the current thresholds shown in \reffig{fig:norm_grad_caffeine}, the time increases by GMS are observed only in the two cases out of 46 cases through \reffig{fig:rmsd_time_test_mols_const1000} and \reffig{fig:time_baker_mols_const300}.

\section{Conclusion} \label{sec:conclusion}
In this work, we propose the scheme to estimate the geometry optimization times at different accuracy levels for a target molecule and the GMS technique that reduces the execution time by dynamically switching multiple methods during geometry optimization. They enable to identify the accuracy levels where geometry optimization will finish in an acceptable time and perform geometry optimization at a selected accuracy level in a shorter time than only using a single method. The evaluation using 46 molecules in total demonstrates that the geometry optimization times at 20 accuracy levels are estimated with a MAPE of 29.5\%, and GMS reduces the execution time by up to 42.7\% without affecting the accuracy of geometry optimization. 

\bibliography{references}

\end{document}